\documentclass[12pt]{article}

\usepackage{xcolor,graphicx}
\usepackage{amssymb,amsmath}
\usepackage{comment}

\usepackage{enumerate}
\usepackage{tikz}
\usepackage[normal,tight,center]{subfigure}
\usepackage{booktabs}
\usepackage{multirow}
\usepackage{slashbox}
\usepackage[Lenny]{fncychap}
\usepackage{caption}
\usepackage{pgfplots}

\usepackage{algorithm}
\usepackage{algpseudocode}
\usepackage{amsmath}
\usepackage{graphics}
\usepackage{epsfig}

\usepackage{empheq}

\def\addlegendimage{\csname pgfplots@addlegendimage\endcsname}
\usetikzlibrary{calc,shadings}

\numberwithin{equation}{section}

\usepackage[colorlinks=true, urlcolor=blue,linkcolor=blue, citecolor=blue]{hyperref}

\DeclareMathAlphabet{\eufrak}{U}{}{}{}
\SetMathAlphabet\eufrak{normal}{U}{euf}{m}{n}
\SetMathAlphabet\eufrak{bold}{U}{euf}{b}{n}

\numberwithin{equation}{section}

\allowdisplaybreaks

 \def\real{{\mathord{\mathbb R}}}
 
 \def\inte{{\mathord{\mathbb N}}}
 
 \def\qu{{\mathord{\mathbb Z}}}

 \def\real{{\mathord{{\rm I\kern-3pt R}}}}        
 
 \def\inte{{\mathord{{\rm I\kern-3pt N}}}}
 \def\sZZ{{\rm Z\kern-.45em{}Z}}
 \def\sQQ{{\kern 0.27em \vrule height1.45ex width0.03em depth0em
           \kern-0.30em \rm Q}}
 \def\qu{{\mathchoice
         {\sQQ}
         {\sQQ}
   {\kern 0.225em \vrule height1.05ex width0.025em depth0em \kern-0.25em \rm Q}
   {\kern 0.180em \vrule height0.78ex width0.020em depth0em \kern-0.20em \rm Q}
         }}
 \def\sGG{{\kern 0.27em \vrule height1.45ex width0.03em depth0em
           \kern-0.30em \rm G}}
 \def\gg{{\mathchoice
         {\sGG}
         {\sGG}
   {\kern 0.225em \vrule height1.05ex width0.025em depth0em \kern-0.25em \rm G}
   {\kern 0.180em \vrule height0.78ex width0.020em depth0em \kern-0.20em \rm G}
         }}

 \newtheorem{prop}{Proposition}[section]

\def\E{\mathop{\hbox{\rm I\kern-0.20em E}}\nolimits}

 \newcounter{hyp}
\title{\huge Option to survive or surrender: carbon asset management and optimization in thermal power enterprises from China}

\author{
Yue Liu
\\
\vspace{-0.1cm} \small
School of Geography, Nanjing Normal University\\
\vspace{-0.1cm}\small
Nanjing 210042, Jiangsu, P.R. China\\
\vspace{-0.1cm} \small
School of Finance and Economics, Jiangsu University\\
\vspace{-0.1cm}\small
Zhenjiang 212013, Jiangsu, P.R. China
\and
Lixin Tian
\thanks{Corresponding author, tianlx@ujs.edu.cn. Supported by the grant from National Natural Science Foundation of China (No: 72004082; 71690242; 11731014), National Key Research and Development Program of China (Grant No. 2020YFA0608601), Jiangsu Natural Science Foundation(BK20180852), Humanities and Social Sciences Foundation of MOE China (18YJA630119), Jiangsu Key Lab for NSLSCS (202006), Project of Philosophy and Social Science Research in Colleges of Jiangsu Province (2020SJA2052).}
\\
\vspace{-0.1cm} \small
Energy Development and Environmental Protection Strategy Research Center
\\
\vspace{-0.1cm}\small
Jiangsu University, Zhenjiang 212013, Jiangsu, P.R. China\\
\vspace{-0.1cm} \small
Energy Interdependency Behavior and Strategy Research Center,
\\
\vspace{-0.1cm}\small
School of Mathematical Science, Nanjing Normal University
\\
\vspace{-0.1cm}\small
Nanjing 210042, Jiangsu, P.R. China
\and
Zhuyun Xie
\\
\vspace{-0.1cm} \small
School of Finance and Economics\\
\vspace{-0.1cm}\small
Jiangsu University, Zhenjiang 212013, Jiangsu, P.R. China\\
\and
Zaili Zhen
\\
\vspace{-0.1cm} \small
Energy Development and Environmental Protection Strategy Research Center\\
\vspace{-0.1cm}\small
Jiangsu University, Zhenjiang 212013, Jiangsu, P.R. China\\
\vspace{-0.1cm}\small
Jiangsu Key Laboratory for Numerical Simulation of Large Scale Complex Systems\\
\vspace{-0.1cm}\small
Nanjing Normal University, Nanjing 210042, Jiangsu, P.R. China\\
\and
Huaping Sun
\\
\vspace{-0.1cm} \small
School of Finance and Economics\\
\vspace{-0.1cm}\small
Jiangsu University, Zhenjiang 212013, Jiangsu, P.R. China
}

\begin{document}
\date{}
\maketitle

\vspace{-1cm}
\newpage
\begin{abstract}
{Carbon emission right allowance is a double-edged sword, one edge is to reduce emission as its original design intention, another edge has in practice slain many less developed coal-consuming enterprises, especially for those in thermal power industry. Partially governed on the hilt in hands of the authority, body of this sword is the prices of carbon emission right. How should the thermal power plants dance on the blade motivates this research. Considering the impact of price fluctuations of carbon emission right allowance, we investigate the operation of Chinese thermal power plant by modeling the decision-making with optimal stopping problem, which is established on the stochastic environment with carbon emission allowance price process simulated by geometric Brownian motion. Under the overall goal of maximizing the ultimate profitability, the optimal stopping indicates the timing of suspend or halt of production, hence the optimal stopping boundary curve implies the edge of life and death with regard to this enterprise. Applying this methodology, real cases of failure and survival of several Chinese representative thermal power plants were analyzed to explore the industry ecotope, which leads to the findings that: 1) The survival environment of existed thermal power plants becomes severer when facing more pressure from the newborn carbon-finance market. 2) Boundaries of survival environment is mainly drawn by the technical improvements for rising the utilization rate of carbon emission. Based on the same optimal stopping model, outlook of this industry is drawn with a demarcation surface defining the vivosphere of thermal power plants with different levels of profitability. This finding provides benchmarks for those enterprises struggling for survival and policy makers scheming better supervision and necessary intervene.
}
\end{abstract}

\baselineskip0.7cm


\section{Introduction}\label{introduction}

\noindent In the global blueprint of carbon emission reduction, China is expected to share a great mission in this campaign \cite{xuan}, and now is the in the great march towards carbon neutrality. From the establishment of carbon emission trading pilot projects in seven provinces and cities in 2011 to the formal launch of carbon emission system in 2017, the volume of China's carbon trading market has expanded rapidly, with a cumulative trading volume of nearly 400 million tons, and the emission reduction of CCER projects has exceeded 300 million tons. However, behind the booming carbon market, there are huge burdens of carbon emission on the shoulders of industries like steel, petrochemical, cement, and thermal power. Since Chinese coal price was increasing in recent years, thermal power industry is under more pressures, more than one hundred of thermal power enterprises went bankrupt. This trend is growing, thanks to the 'Big up small down' scheme, that is, to put forward the big generator sets and oppress the small one. As well hyped, more giant units were under construction, such like Guangdong Huadian Fengsheng Shantou power plant of 2*660MW launched formally from Oct. 2020, and Zhanjiang Jingxin Donghai power plant of 2*600MW, planned to put into operation in 2022. It is believed that, although under huge pressure of energy transformation, thermal power still accounts for a constantly large portion for power generation, as shown in the Figure \ref{f0} below.
\begin{figure}[htbp]
\vskip-0.9cm
\hskip -3.0cm
\centering
\includegraphics[height=1.3 \textwidth,width=1.2\textwidth]{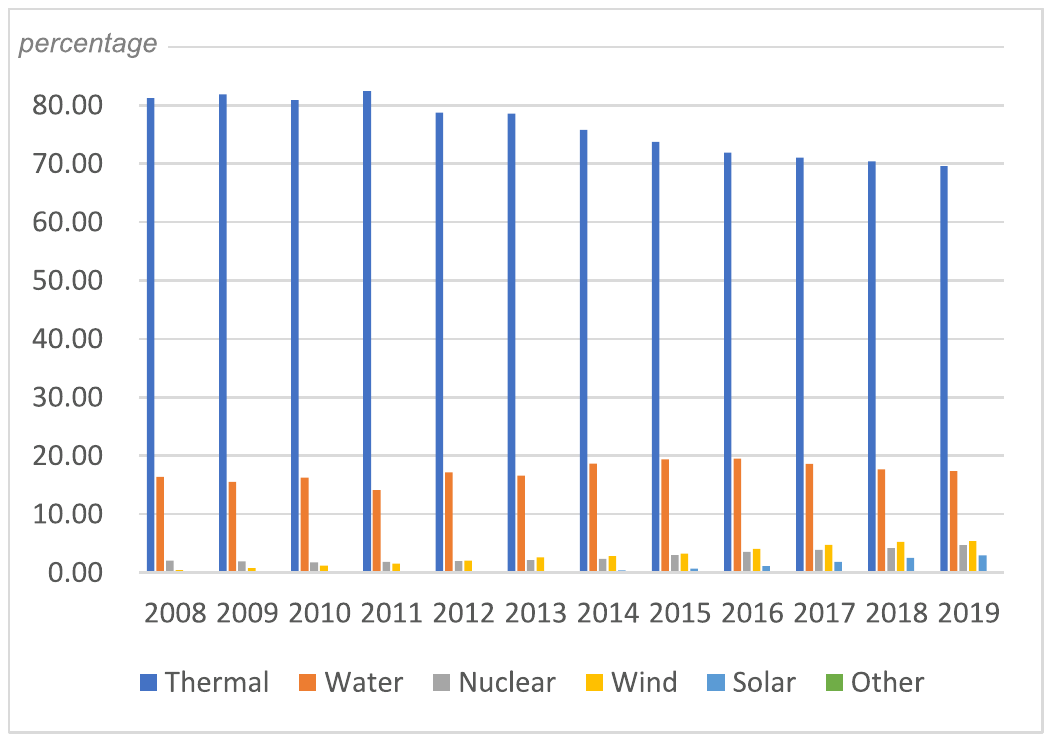}
\vskip-12.8668cm
	\caption{\small{Structure of generating capacity (\%)}}
	\label{f0}
\end{figure}
In the shadow of the newborn giants, traditional thermal power plants are struggling. As pointed by Shumin Zhang, the chief economist of Guodian group (top 5 in this industry in China), 'thermal power plants are as difficult as steel enterprises and coal enterprises, and may go bankrupt in large numbers. They are not forced to close down by the government, but are unable to operate hence close down'. So what are their problems of operation? In this paper, we investigate the operation and survival problem of thermal power enterprises from the perspective of profit achievable under the impact of carbon emission allowance market.\\

\noindent Carbon emission allowance is the core asset in carbon finance market, it transfers the government's restraint in carbon emission to the enterprises in form of its prices. Hence for those companies, carbon asset operation becomes a crucial aspect of enterprise operation. To describe the strategy and actions during the carbon asset operation, Markov decision process is applied to simulate the decision-making as in \cite{li} for energy storage system, \cite{zhuang} for management of greenhouses, \cite{zhang} for optimizing energy conversion and \cite{xiong} for micro-grid power optimal control. Paralleling to the approach of Markov decision process, optimal stopping model is also usually used to characterize the timing of decision and operation, especially for those financial environments with continuous dynamic of price process, see \cite{ferrari} for asset trading strategy and \cite{wangh}, \cite{yang},  \cite{bayer} for option pricing. The former has its advantage of describing the decision-making for all points in the time horizon, while the later seeks one or limited several (multiple optimal stopping times) timings for some specific action and it handles both discrete and continuous environment. In the operation modelling of this paper, we mainly focus on the time to suspend or halt production under pressure of profitability and the decision is under consideration of time-continuous price dynamic, hence optimal stopping problem is established after modeling the price process with geometric Brownian motion, which is commonly used to simulate the variable on continuous time horizon, as applied in \cite{ramos} for price forecasting, \cite{hoyyi}, \cite{abensur} for stock trading, \cite{long} for risk analysis and hedging.
\\

\noindent There have been intensive researches on the topics of carbon emission allowance, carbon capital operation and thermal power plants. Thermal power is one of the most signification industries contributing to carbon reduction, especially the clear power resources like solar thermal power, which has been developed fast and taken an increasing share of power industry, as shown by \cite{behar} and \cite{reddy}. More researches were about its technical specialties, like its performance (see \cite{behar1} and \cite{bishoyi}) and parabolic trough (see \cite{aqachmar}-\cite{salazar}). 
Besides, its technical and economic potentials are widely and repeatedly analyzed, as in \cite{adibhatla} and \cite{purohit}. Compared with those above mentioned with researches on traditional thermal power like \cite{helisto} and \cite{partridge}, concerns for the coal powered plants are emerging since it appears that the shifting of weight was speeding up in the thermal power industry, which arguments are supported by perspectives from \cite{bai}. Specifically when taking a close check over this industry in China, they tend to believe that survival environment of traditional thermal power plants is shrinking for several reasons, one is increasing in the overall cost (see \cite{huaping} for more information), influenced mainly by the prices of stream coal and carbon emission allowances, another is the restrained electricity selling price, which is generally not very market-orientated, but more governed by the authority. Under pressures from both ends, hundreds of plants have been squeezed out, survivals may have their distinctive advances in emission reduction technology (as shown by \cite{mahmoudi} and \cite{dmitrienko}), cost control (such as those in \cite{schill} and \cite{eser}), efficiency improvement (as proposed by \cite{haseli} and \cite{jimenez}), or operation management optimizations (refer to \cite{helisto2} and \cite{wangl}). Integrating all aspects from technical escalation, operation management and market impact, we aim to investigate the survival environment of Chinese thermal power plants and decode the processes of their failures and survivals.\\

\noindent This paper contributes in both theoretical and practical aspects. It introduces the optimal stopping model to the management of carbon asset, particularly, the novelty relies on its application to describe the enterprise decision under the impact of carbon price. Besides, it reveals the profitability required for survival at cost of one unit of carbon emission, based on which it uncovers the underlying process of the collapse of a thermal power plant under the pressure of carbon price as well as other plants' tough ways of survival hence illustrates some significant managemental implications.\\

\noindent We proceed as follows. In Section \ref{formulation}, price process of carbon emission allowance is modeled by geometric Brownian motion and the optimization of operation management with carbon asset is modeled as an optimal stopping problem, which is analyzed and algorithmically solved in Section \ref{solving}. By solving the relevant optimal stopping problems, shutdown of Shajiao B power plant in 2019 and struggling for life extension of Wushashan Power plant during 2017-2019 are investigated in Section \ref{shut} and \ref{surv} respectively. In line with the above case studies and by the same methodology, Section \ref{outlook} presents the outlook of the survival environment of Chinese thermal power industry. Finally, Section \ref{conclu} concludes to the theoretical and case analysis before illustrating some management implications and suggestions.


\section{Formulation}\label{formulation}
To formulate the core problem in the scenarios of operation management with carbon asset of a coal-powered plant, we define several variables first. Let $M\in\real_+$ denote the averaged daily emission, $(Y_t)_{t\in[0,T]}$ denotes the price of carbon emission right at time $t\in[0,T]$, where $[0,T]$ is the production cycle. During this cycle, the averaged profit achievable at cost of a unit of carbon emission is denoted as $P\in\real_+$.
\\

\noindent Hence if the price of carbon emission right $Y_t$ is always well above the profit $P$, this enterprise will consider a reduction or halt of output. This decision might not be taken immediately once the price $Y_t$ excesses $P$ since some fluctuation of price may create price peaks occasionally and influence little on the judgement of the long-term behavior of price process. However, in view of the probabilistically expectation of the comparison between price $Y_t$ and profit $P$, if maintenance of production for the whole time horizon $[0,T]$ will finally yield inferior results, towards the target of ultimate maximization of total net profit, this enterprise will optimize an halting time to quit this process of loss and we assume it will finally sell out the remaining carbon emission right at time $T$. This stopping time is denoted as $\tau\in[0,T]$. Obviously, an extreme case that $\tau=T$ illustrates that production sustains during this whole cycle and the enterprise is in no need of such intervene. In this case, total carbon emission during the production cycle is $MT$, and the total profit becomes $MTP$. For the general case with consideration of production halt, total profit $R(\tau)$ will be expressed in the reward formula without excluding the extreme case $\tau=T$. As follows, the reward function of optimal stopping time is defined:
\begin{equation}\label{rtau}
R(\tau):=MP\tau+MY_T(T-\tau).
\end{equation}
To simulate the carbon emission right price process $(Y_t)_{t\in[0,T]}$, we simply apply a geometric brownian motion as many existed researches did to capture the price dynamic:
\begin{equation}\label{yt}
dY_t=\mu Y_tdt+\sigma Y_t dB_t, \qquad t\in[0,T],
\end{equation}
where $\mu\in\real_+$ is the drift factor and $\sigma\in(0,\infty)$ is the volatility factor, $(B_t)_{t\in\real_+}$ is a standard Brownian motion, we denote ${\cal F}_t$ as the filtration generated by $\sigma\{B_s; s\in[0,t]\}$. The solution to the stochastic differential equation \eqref{yt} is given by
\begin{equation}\label{yt2}
Y_t=Y_0\exp\left((\mu-\frac{\sigma^2}{2})t+\sigma B_t\right),
\end{equation}
for more details, we refer to \cite{oksendal}.
\\

Since enterprises are aim to maximize the final profit, we search for an optimal stopping time $\tau^*$ among all possible stopping times $\tau\in[0,T]$. This stopping time $\tau^*$ is theoretically optimal in a sense of achieving the ultimate profit maximization, which is in form of the following optimal stopping problem:
\begin{equation}\label{m}
V=\sup\limits_{\tau\in[0,T]}\E[R(\tau)],
\end{equation}
of which the solution $\tau^*$ satisfying the equation that $V=\E[R(\tau^*)]$.
For further investigate this optimal stopping problem, we define the value function $V(t,y)$ as follows for $t\in[0,T]$ and $y\in\real_+$:
\begin{equation}\label{m2}
V(t,y)=\sup\limits_{\tau\in[t,T]}\E[R(\tau)|Y_t=y].
\end{equation}

To justify the existence of the optimal stopping time according to \cite{book1}, we need to define a function $G(t,y)$ that
\begin{equation}\label{g}
G(t,y)=\E[R(t)|Y_t=y]
\end{equation}
\noindent and check some boundedness and smoothness conditions as follows.
\begin{enumerate}[a)]
\item $G(t,y)$ is lower semicontinuous with respect to $y$. This is easily checked by expressing $G(t,y)$ after combining \eqref{rtau}, \eqref{yt2} and \eqref{g}:
\begin{align}\label{g2}
&G(t,y)=\E[MPt+Mye^{(\mu-\frac{\sigma^2}{2})(T-t)+\sigma (B_T-B_t)}(T-t)]\nonumber\\
&=MPt+Mye^{\mu(T-t)}(T-t).
\end{align}
\item $V(t,y)$ is upper semicontinuous with respect to $y$, which is checked by similar approaches with \eqref{rtau}, \eqref{yt2} and \eqref{m2}.
\item $G(t,y)<\infty$ by checking the expression \eqref{g2}.
\end{enumerate}
Following the standard arguments as in \cite{book1}, \cite{mine19}, \cite{mine18}, we define a stopping set $D$ by $D:=\{(t,y)\,|\, V(t,y)=G(t,y)\}$ and its complementary set $G$ (called as continuation set) is given by $C:=\{(t,y)\,|\, V(t,y)>G(t,y)\}$. With the continuity of $V$ and $G$, it is easily checked that $D$ is a closed set and $C$ is an open set. Besides, we can define the boundary of $D$ and $C$ as ${\cal B}:=D\cap {\bar C}$, where ${\bar C}$ is the closure of set $C$. This boundary, called as free boundary, will be shown in Proposition \ref{propa} that, it is in form of a curve determined by a one-to-one mapping of $t\rightarrow y$.

\section{Solution to the optimal stopping problem}\label{solving}
Substituting \eqref{yt2} into \eqref{rtau} and \eqref{m2} yields that
\begin{equation}\label{m3}
V(t,y)=\sup\limits_{\tau\in[t,T]}\E[MP\tau+yM\exp\left((\mu-\frac{\sigma^2}{2})(T-t)+\sigma (B_T-B_t)\right)(T-\tau)|Y_t=y].
\end{equation}
By \eqref{m3} and applying strong Markovian property, we see that
\begin{align*}
&V(t,y)=\sup\limits_{\tau\in[t,T]}\E[\E[MP\tau+yMe^{(\mu-\frac{\sigma^2}{2})(T-t)+\sigma (B_T-B_t)}(T-\tau)|{\cal F}_\tau]]\nonumber\\
&=\sup\limits_{\tau\in[t,T]}\E[\E[MP\tau+yMe^{(\mu-\frac{\sigma^2}{2})(T-t)+\frac{\sigma^2}{2}(T-\tau)+\sigma (B_\tau-B_t)}(T-\tau)|{\cal F}_\tau]]\nonumber\\
&=\sup\limits_{\tau\in[t,T]}\E[MP\tau+yMe^{\mu(T-t)-\frac{\sigma^2}{2}(\tau-t)+\sigma (B_\tau-B_t)}(T-\tau)]\nonumber\\
&=\sup\limits_{\tau\in[0,T-t]}\E[MP\tau+yMe^{\mu(T-t)-\frac{\sigma^2}{2}\tau+\sigma B_\tau}(T-\tau-t)]+MPt.
\end{align*}
Applying the result of exponential martingale to the formula above and by \eqref{g2} of $G(t,y)$, it follows that
\begin{align}\label{m4}
&V(t,y)=\sup\limits_{\tau\in[0,T-t]}\E[P\tau- y\tau e^{\mu(T-t)-\frac{\sigma^2}{2}\tau+\sigma B_\tau}]M+MPt+yM(T-t)e^{\mu (T-t)}\nonumber\\
&=\sup\limits_{\tau\in[0,T-t]}\E[P\tau- y\tau e^{\mu(T-t)-\frac{\sigma^2}{2}\tau+\sigma B_\tau}]M+G(t,y).
\end{align}
As revealed by \eqref{m4}, $V(t,y)-G(t,y)$ is decreasing in $y$ for any $t\in[0,T]$ and $y\in\real_+$. This property implies that given $V(s,y)=G(s,y)$ for certain $s\in[0,T]$, this equality holds for any $t\in[s,T]$ and a proposition follows:
\begin{prop}\label{propa}
If stopping should be taken under observation of carbon emission right price $Y_t$ at time $t\in[0,T]$, stopping is optimal for any higher prices of $Y_t$ at time $t$. Besides, stopping set $D$ and $C$ is hence spitted by a curve $\{(t,y): y=b(t)\}$, which becomes the free boundary ${\cal B}$.
\end{prop}
For any $s\in[0,T]$ and by the definition of optimal stopping time, by \eqref{m4}, we see that
\begin{align*}
V(t,y)\geq \E[Ps- ys e^{\mu(T-t)-\frac{\sigma^2}{2}s+\sigma B_s}]M+G(t,y)=(P- ye^{\mu(T-t)})Ms+G(t,y).
\end{align*}
Hence we conclude that:
\begin{prop}\label{propb}
Suppose the price of carbon emission right $Y_t$ at time $t\in[0,T]$ is below $Pe^{-\mu(T-t)}$, where $P\in\real_+$ is the averaged profit achievable at cost of a unit of carbon emission, this enterprise will maintain the state of production. Therefore, $Pe^{-\mu(T-t)}<b(t)$ for any $t\in[0,T]$.
\end{prop}
In view of Proposition \ref{propa} and \ref{propb}, a rough draft is drawn as Figure \ref{f1} below. The red curve $y=b(t)$ denotes the free boundary, separating the stopping set $D$ and continuation set $C$, and the blue curve $y=Pe^{-\mu(T-t)}$ (for a positive $\mu$) below the red one and inside the continuation set $C$ is a lower bounder line of stopping set $D$.

\begin{figure}[htbp]
\centering
\includegraphics[height=0.8 \textwidth,width=0.68\textwidth]{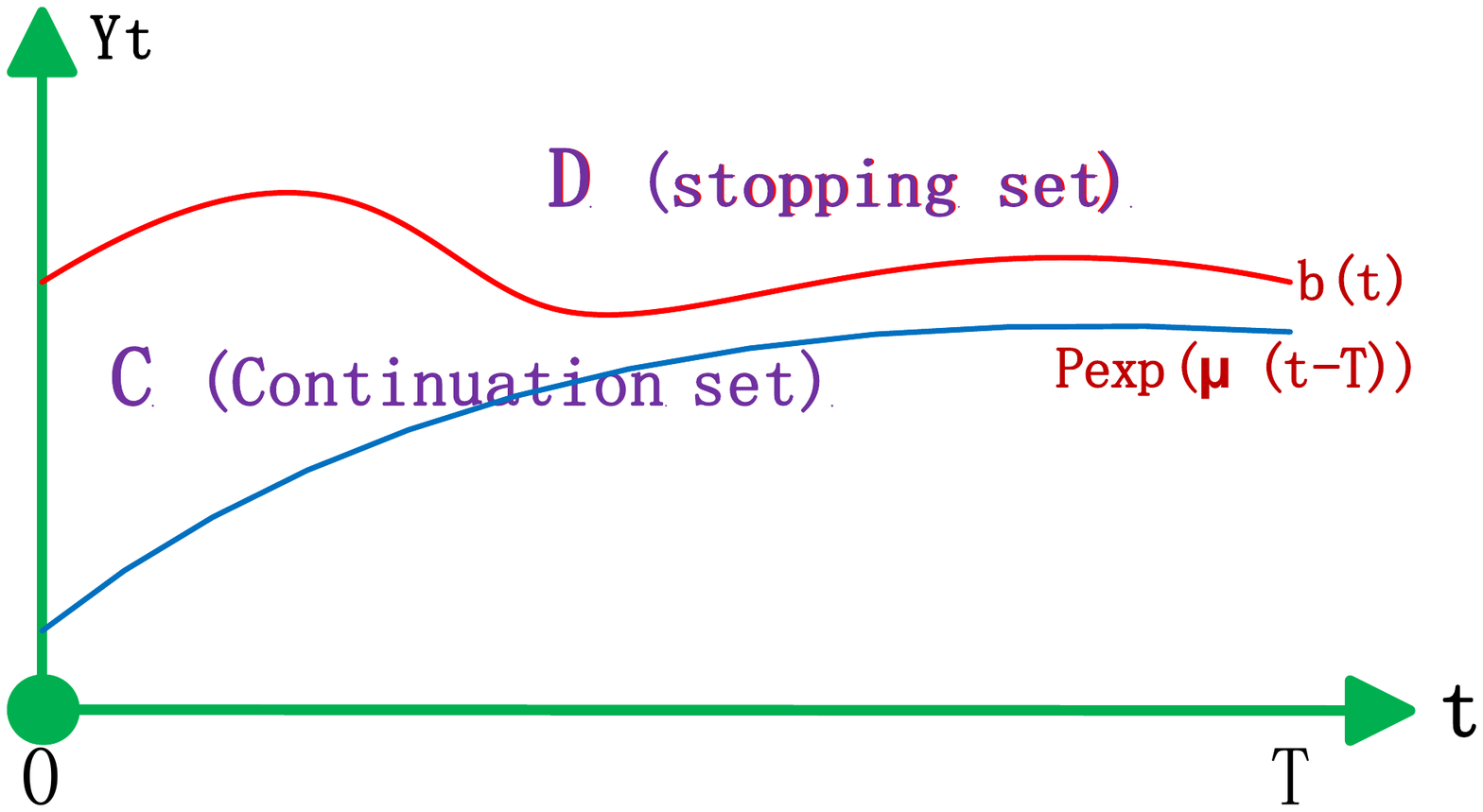}
\vskip-7.668cm
	\caption{\small{Stopping set and free boundary}}
	\label{f1}
\end{figure}
\noindent To solve this optimal stopping problem, it is most preferred to have an explicit expression of $b(t)$ for all $t\in[0,T]$. However, this is not achievable in most cases. Particularly, for the optimal stopping problem \eqref{m3} we considered, there is no standard approach to solve it by stochastic analysis and to conclude with a closed-from expression. Instead, with all parameters well collected, we can design a backward algorithm to compute $b(t)$ numerically. The basic methodology is to calculate the value function after discretization of time horizon. Particularly, $[0,T]$ is discretized into a sequence $\{t_0,\ldots,t_n\}$ denoted as $S(\delta)$ with $t_0=0$, $t_n=T$, $t_{i+1}-t_i=\delta>0$. To solve the optimal stopping problem in \eqref{m4}, we consider the discrete optimal stopping problem of $\sup\limits_{\tau\in S(\delta)\cap[0,T-t]}\E[P\tau- y\tau e^{\mu(T-t)-\frac{\sigma^2}{2}\tau+\sigma B_\tau}]$ denoted as $U(t,y)$ for $t\in S(\delta)$, $x\in\real_+$. The recursion relation is obtained in the following formula for $i\in\{0,\ldots,n-1\}$ that
\begin{align}\label{rec}
&U(t_i,y)=\sup\limits_{\tau\in S(\delta)\cap[0,T-t_i]}\E[P\tau- y\tau e^{\mu(T-t_i)-\frac{\sigma^2}{2}\tau+\sigma B_\tau}]\nonumber\\
&=\max\left(0,\sup\limits_{\tau\in S(\delta)\cap[\delta,T-t_i]}\E[P\tau- y\tau e^{\mu(T-t_i)-\frac{\sigma^2}{2}\tau+\sigma B_\tau}]\right)\nonumber\\
&=\max\left(0,\sup\limits_{\tau\in S(\delta)\cap[\delta,T-t_i]}\E[P\tau- y\tau e^{\mu(T-t_i)-\frac{\sigma^2}{2}\tau+\sigma B_\tau}]\right)\nonumber\\
&=\max\left(0,\sup\limits_{\tau\in S(\delta)\cap[0,T-t_{i+1}]}\E\left[(\tau+\delta)\left(P- y e^{\mu(T-t_i)-\frac{\sigma^2}{2}(\tau+\delta)+\sigma B_{\tau+\delta}}\right)\right]\right)\nonumber\\
&=\max\left(0,\sup\limits_{\tau\in S(\delta)\cap[0,T-t_{i+1}]}\E\left[\tau\left(P- y e^{\mu(T-t_i)-\frac{\sigma^2}{2}(\tau+\delta)+\sigma B_{\tau+\delta}}\right)\right]\right.\nonumber\\
&\quad \left.+\delta(P- y e^{\mu(T-t_i)})\right)\nonumber\\
&=\max\left(0,\E\left[\sup\limits_{\tau\in S(\delta)\cap[0,T-t_{i+1}]}\E\left[\tau\left(P- y e^{\mu(T-t_i)-\frac{\sigma^2}{2}(\tau+\delta)+\sigma B_{\tau+\delta}}\right)\,|\,e^{\sigma(B_{\tau+\delta}-B_\tau)}\right]\right]\right.\nonumber\\
&\quad \left.+\delta(P- y e^{\mu(T-t_i)})\right)\nonumber\\
&=\max\left(0,\E\left[U(t_{i+1},ye^{\mu\delta-\frac{\sigma^2}{2}\delta+\sigma(B_{\tau+\delta}-B_\tau)})\right]+\delta(P- y e^{\mu(T-t_i)})\right).
\end{align}
\noindent Accompanied with this formula, we design the algorithm whose pseudocode is attached in the Appendix.

\section{Shutdown of Shajiao B power plant}\label{shut}
Shajiao B power plant was one of China legends, and famous for remarkable records: it was the first Chinese power plant constructed partially by foreign capital and in form of BOT model, its construction took less than 2 years noted as the fastest speed in 1980's, moreover, the whole base of Shajiao power plant exceeds all others in southern China. Overall, Shajiao power plant was always regarded as one of the most significant landmarks of China reform and opening-up. However, this legend has been terminated. After its 32 years of glory, Shajiao B power plant was shut down at 13th November, 2019. Many reasons accounts for this, but besides the environmental demands, despairing unprofitable operation further pulls it down. It will be illustrated by the optimal stopping model established in Section \ref{formulation} and solved by the proposed method in Section \ref{solving}.
\\

\noindent To investigate the profitability during Shajiao B's dying period, we collect all sorts of data from 2018 to 2019. Since Shajiao power was registered in Shenzhen emissions exchange and mainly traded on SZA-2014 emission right allowance which is issued in 2014 and widely transacted with high volume till now, we consider the SZA-2014 during this period, as shown in Figure \ref{sza} below.
\begin{figure}[htbp]
\vskip-0.82868cm
\centering
\includegraphics[height=0.88 \textwidth,width=0.8\textwidth]{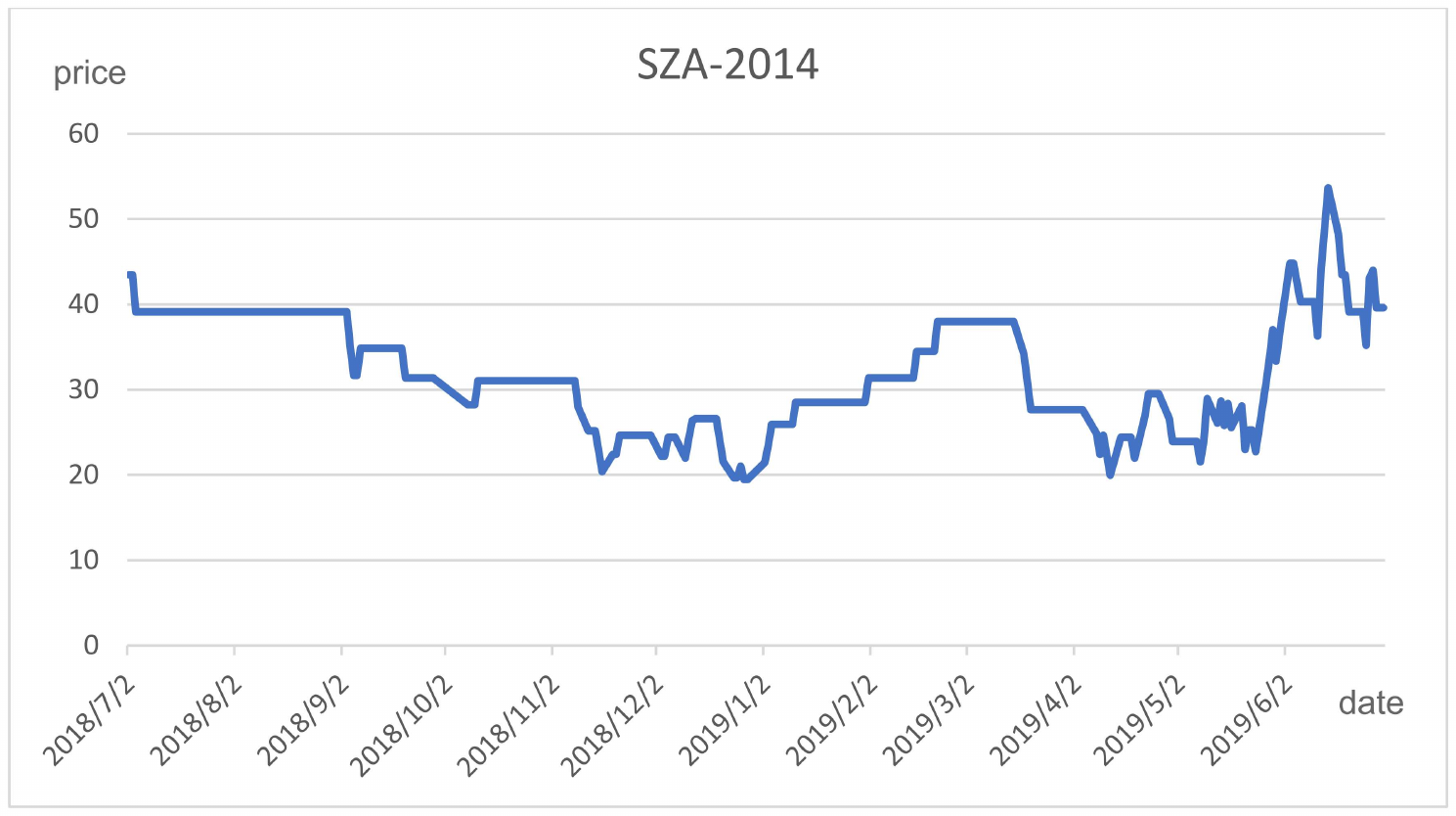}
\vskip-8.368cm
	\caption{\small{Prize dynamics of SZA-2014 from Jul. 2018 to Jun. 2019 in Shenzhen emissions exchange}}
	\label{sza}
\end{figure}

\noindent Shajiao B power plant was founded in 1987 by Guangdong Guanghe Power Co., Ltd. held by Guangdong Yuedian Group Co., Ltd. ($60\%$), China Resources Power Co., Ltd. ($30\%$) and China Huaneng Group Co., Ltd. ($5\%$). The installed capacities of both unit 1 and unit 2 of power plant B are 350MW. After site investigation and survey, we find that during the year 2018, the total coal consumption is about 1.88 million tons, with about 3.52 million tons of carbon dioxide emissions. The Annual revenue is approximately 51.7 million CNY. Averaged over 246 trading days of SZA in 2018, daily emission is about 0.0143 million tons. Hence we let $M=0.0143$ and $P=14.7$, which is the averaged revenue at the cost of one unit of carbon emission. Without full information of year 2019 when making prediction and decisions in year 2019, we assume that $M=0.0143$ and $P=14.7$ remains the same during the year 2019 since there were no significant technical improvement observed by field investigation.
\\

\noindent For parameter estimation of the geometric Brownian motion \eqref{yt} of emission right allowance price of 2018, $\sigma$ and $\mu$ are determined by the price data following the approaches below:
$1)$ All trading data in this period are collected, let $T=246$ (trading days), prices are listed as ${Y_i}_{i\in{0,\ldots,n}}$; $2)$ Compute the return $r_i$ for $i\in\{1,\ldots,n\}$ by $r_i=\log(Y_{i})-\log(Y_{i-1})$; $3)$ Compute the historical volatility $\sigma_h$ and historical drift $\mu_h$ according to \cite{hull} and \cite{mine19} below:
\begin{equation}\label{sigmah}
\mu_h=\bar{r}:=\frac{1}{T}\sum\limits_{i=1}^T r_i,\quad \sigma_h=\sqrt{\frac{1}{T-1}\sum\limits_{i=1}^T(r_i-\bar{r})^2},
\end{equation}
which are the commonly-used estimators of $\mu$ and $\sigma$ in \eqref{yt}. Decision made in 2019 is based on all visible information by that time, hence the price model is established based on in-sample data, namely the daily quotations in 2018. Feeding the above formula \eqref{sigmah} with data of year 2018 immediately yields the parameters for \eqref{yt}, that is, $\mu=-0.0020$, $\sigma=0.0603$. Starting with $Y_0=21.43$, geometric Brownian motion \eqref{yt} simulates the price process, which is applied in the model \eqref{m4}. Hence to solve the optimal stopping problem \eqref{m4}, all parameters are ready as shown by Table 1.
\begin{center}
\begin{table}
\footnotesize
\label{table1}
\centering
\begin{tabular}{cccccccc}
\toprule
$Y_0$ & $\mu$ & $\sigma$ & $T$ &$P$ & $M$ \\
\midrule
21.43&-0.0020&0.0603&  246 & 14.7 & 0.014\\
\bottomrule
\end{tabular} \caption{Parameters of the optimal stopping problem for analyzing the case of Shajiao B power plant in 2019}
\end{table}
\end{center}
\vskip-1.18cm
Next, implementing a backward recursive algorithm whose pseudocode is attached in the Appendix to solve the optimal stopping problem, we obtain the boundary and show the actual operation with consideration of the carbon emission price in Figure \ref{f2} below.
\begin{figure}[htbp]
\vskip-1.38cm
\centering
\includegraphics[height=1.328 \textwidth,width=0.98\textwidth]{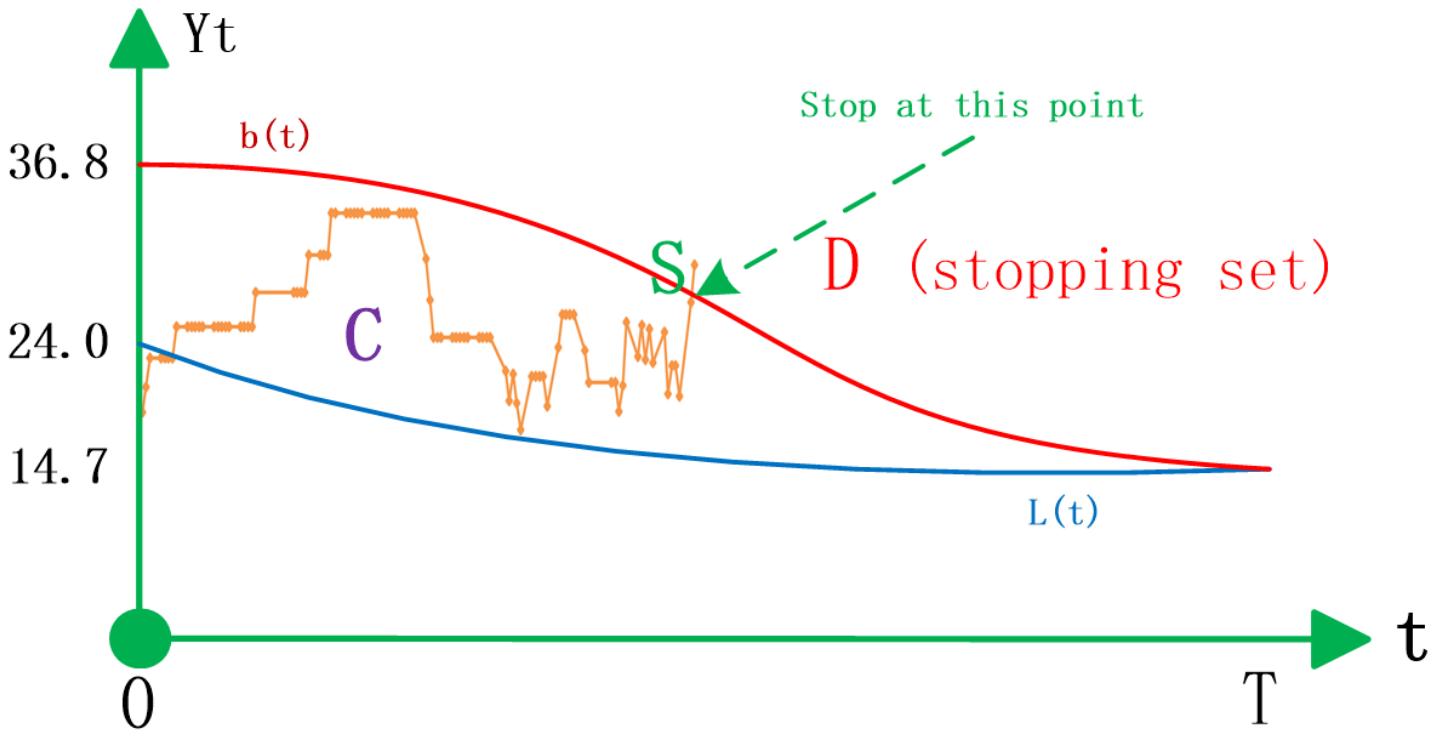}
\vskip-13.68668cm
	\caption{\small{Optimal stopping boundary and actual operation of Shajiao B in 2019}}
	\label{f2}
\end{figure}
To interpret Figure \eqref{f2}, we introduce the curves and nodes in this graph. The blue line $L(t)$ denotes a lower bounder of stopping set $D$, as also plotted in Figure \ref{f1}. But since $\mu<0$ in this case, $L(t)$ is decreasing unlike that in Figure \ref{f1}. Above the lower bounder line $L(t)$ we see the free boundary line $b(t)$ separating the stopping set $D$ and continuation set $C$ below. $b(t)$ is also decreasing although it is not technically ready to investigate its monotony in general cases. An intersection of the two curves is $(T,P)$, where $T=246$ and $P=14.7$. At this point, there is no remaining time slot for hesitation. At the other end, $b(t)$ starts at 36.8, which is the lowest value of $y$ satisfying $V(0,y)=G(0,y)$. Besides the two curves, an orange broken line records the market price of carbon emission (SZA-2014) during the first half year of 2019, marked nodes on this line denotes the dairy price with nonzero volume. Obviously, in the practice of modelling, this market price curve is plotted with out-of-sample data.
\\

\noindent During the operation of the power plant, this orange broken line should be updated on each trading day by adding new points and extending the line on this graph with the curve $b(t)$ depicted in advance. Meanwhile, the plant manager should observe every day to check whether the price curve hits the free boundary $b(t)$. In the scenario recorded in Figure \ref{f2}, the price curve breaks the boundary $b(t)$ at the point $(96,33.66)$ marked as point $S$ in the graph. This breakthrough suggests an stopping of the current process with perspectives of ultimate profit during the whole time horizon. Therefore, as the facts we see, Shajiao B power plant terminated two main units and never restarted them till the overall closing of the plant at 13th Nov. 2019.

\section{Survivors' way out}\label{surv}
The consistence between the case of Shajia B power plant and our model is not an isolate coincidence, and not all power plants were following a similar free boundary of stopping, shown as a red curve in Figure \ref{f2}, which would finally be broken by the uprising carbon emission rights price process during year 2019. However, there exists also many survivors flying in the face of the increasingly stringent carbon regulation, upward market of carbon emission rights, as well as the uprising price of coal. In this section, we aim to see the difference of carbon asset management between the failure and survivor, shown by the proposed model and their free boundaries of stopping.
\\

\noindent There are five top groups in the industry of thermal power in China, namely China Huaneng Group, China Datang Corporation, China Huadian Corporation, China Guodian Corporation, China Power Investment Group, who are actually controlling or partially controlling over 80 percent of power plants in China. The above mentioned Shajiao B power plant was only 5$\%$ held by China Huaneng Group, but more big power plants are widely controlled with the five giants. So in this section we consider
Zhejiang Datang Wushashan Power Generation Co., Ltd for investigation. To explore its mystique, we focus on its cost of a unit of carbon emission formerly denoted as $P\in\real_+$ and technical innovation accounting for these changes. Indeed, among its many rounds of technical upgrading in recent years, two of them are more notable. One is achieved in autumn of 2017, resulting in energy saving in denitration process, sulfur trioxide dust control of coal burning, high efficiency dust removal and intelligent control of coal fired units. Another round of technical escalation we observed is in winter of 2019, by successively developing and applying the AGC and steam temperature intelligent control system, effectively improving the automatic control performance of the unit, and with the "two detailed rules" of the unit, Wushashan power plant has achieved substantial profits. Hence we collect data regarding these two events.
\\

\noindent Wushashan power plant has four units of coal fired generating sets, of which the total capacities reaches 2.4 million KW. After site investigation and survey, we find that during the year 2016, the total coal consumption is about 6.08 million tons, with about 11.68 million tons of carbon dioxide emissions. The Annual revenue is approximately 156.2 million CNY. Averaged over 243 sample days of 2016, daily emission is about 0.048 million tons. Hence we let $M=0.048$ and $P=14.5$, which is the averaged revenue at the cost of one unit of carbon emission. Implementation of technical improvement in Jul. 2017 dramatically changed the value $P$ and $M$, we denote the updated values by ${\tilde P}$ and ${\tilde M}$. From the results that ${\tilde P}= 17.2$ and ${\tilde M}=0.041$, we see that the utilization of carbon is improved. Before analyzing the case of 2017, the carbon emission allowance price dynamic is modeled based on the observations of prices in 2016. Since Wushashan power plant is registered in Shanghai Environment and Energy Exchange, it trades on SHEA (Shanghai Emission Allowance), see Figure \ref{shea} below.
\\

\begin{figure}[htbp]
\vskip-1.8868cm
\centering
\includegraphics[height=1.08 \textwidth,width=0.98\textwidth]{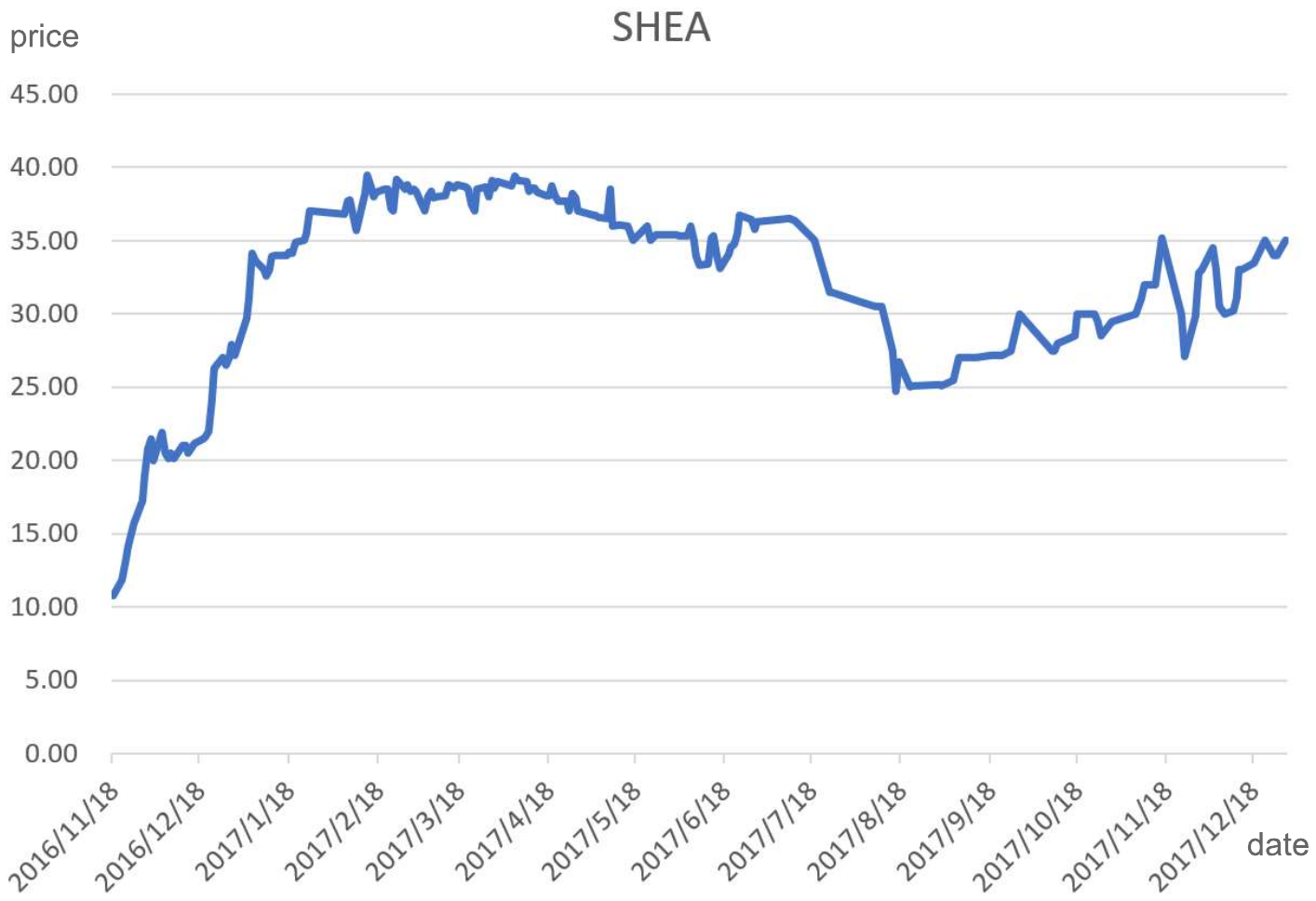}
\vskip-9.08368cm
	\caption{\small{Prize dynamics of SHEA at Shanghai Environment and Energy Exchange during year 2016-2017}}
	\label{shea}
\end{figure}

\noindent To analysis the decision making in 2017 with all visible information by that time, price model should be established based on in-sample data, namely the daily SHEA quotations in the first half year of 2017. Feeding the formula \eqref{sigmah} with data of first half year of 2017 (because SHEA opens at 18th Nov. 2016, to model the market of second half of 2017, we use the data of first half of 2017) immediately yields the parameters for \eqref{yt}, that is, $\mu=-0.0019$, $\sigma=0.0238$. SHEA has much less trading days than that of SZA, it has 49 trading days in the second half year of 2017. Starting with $Y_0=36.5$ (SHEA quotation of July. 10th, 2017), geometric Brownian motion \eqref{yt} simulates the price process, which is applied in the model \eqref{m4} with parameters shown by Table 2 below.

\begin{center}
\begin{table}
\footnotesize
\label{table2}
\centering
\begin{tabular}{ccccccccccc}
\toprule
$Y_0$ & $\mu$ & $\sigma$ & $T$ &$P$ & $M$ &${\tilde P}$ & ${\tilde M}$ \\
\midrule
36.50& -0.0019 &0.0238&  49 & 14.5 & 0.048 & 17.2 & 0.041\\
\bottomrule
\end{tabular} \caption{Parameters of the optimal stopping problem for analyzing the case of Wushashan power plant in 2017}
\end{table}
\end{center}
\vskip-1.18cm

\begin{figure}[htbp]
\vskip-1.0868cm
\centering
\includegraphics[height=1.08 \textwidth,width=0.98\textwidth]{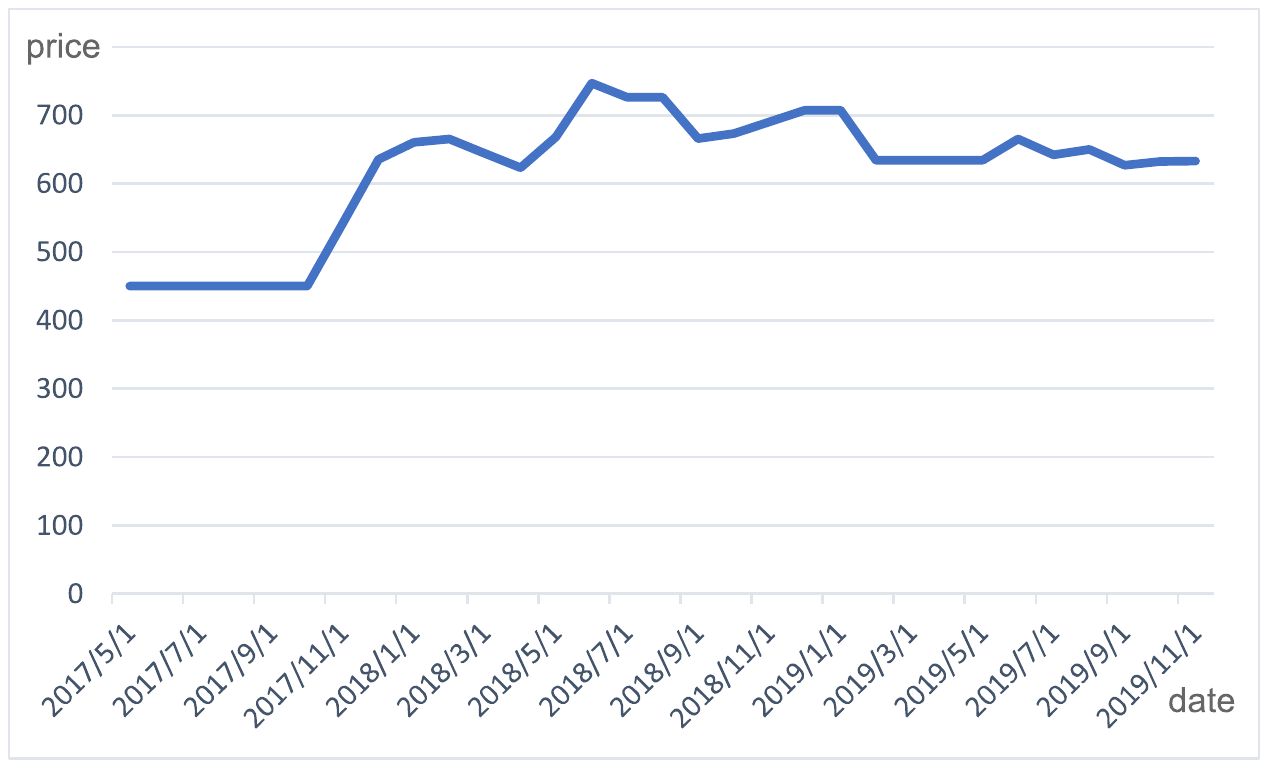}
\vskip-10.68368cm
	\caption{\small{Prices of steam coal Q5500 during year 2017-2019}}
	\label{coal}
\end{figure}

\noindent For another case of Wushashan power plant in Nov. 2019, it was mainly about the technical improvement of steam temperature intelligent control system and automatic control performance of the unit. During the two years, domestic coal price is rising by about 40 percent, as shown in Figure \ref{coal}, where the quotations of Q5500 (calorific value is 5500 kcal) steam coal before Mar. 2018 are most from Jinchen city and those after Mar. 2018 are all from Ningbo port. This intentional selection is in line with the actual transaction Wushashan power plant made during the previous years.\\

\noindent The uprising coal price dramatically brings down the revenue denoted as $P$, there were several technical escalations during these two years, yet the revenue raises little after deducting the research and development expenditure. But the technical improvement in Nov. 2019 is comparatively significant. Similar approaches of investigation and model parameter estimation yield the Table 3 as follows, where the model parameters are estimated with SHEA quotations of year 2018 and first half year of 2019 applied as in-sample data and the SHEA prices during 2018 and 2019 are plotted in Figure \ref{price1819} below.

\begin{figure}[htbp]
\vskip-1.2868cm
\centering
\includegraphics[height=1.08 \textwidth,width=0.98\textwidth]{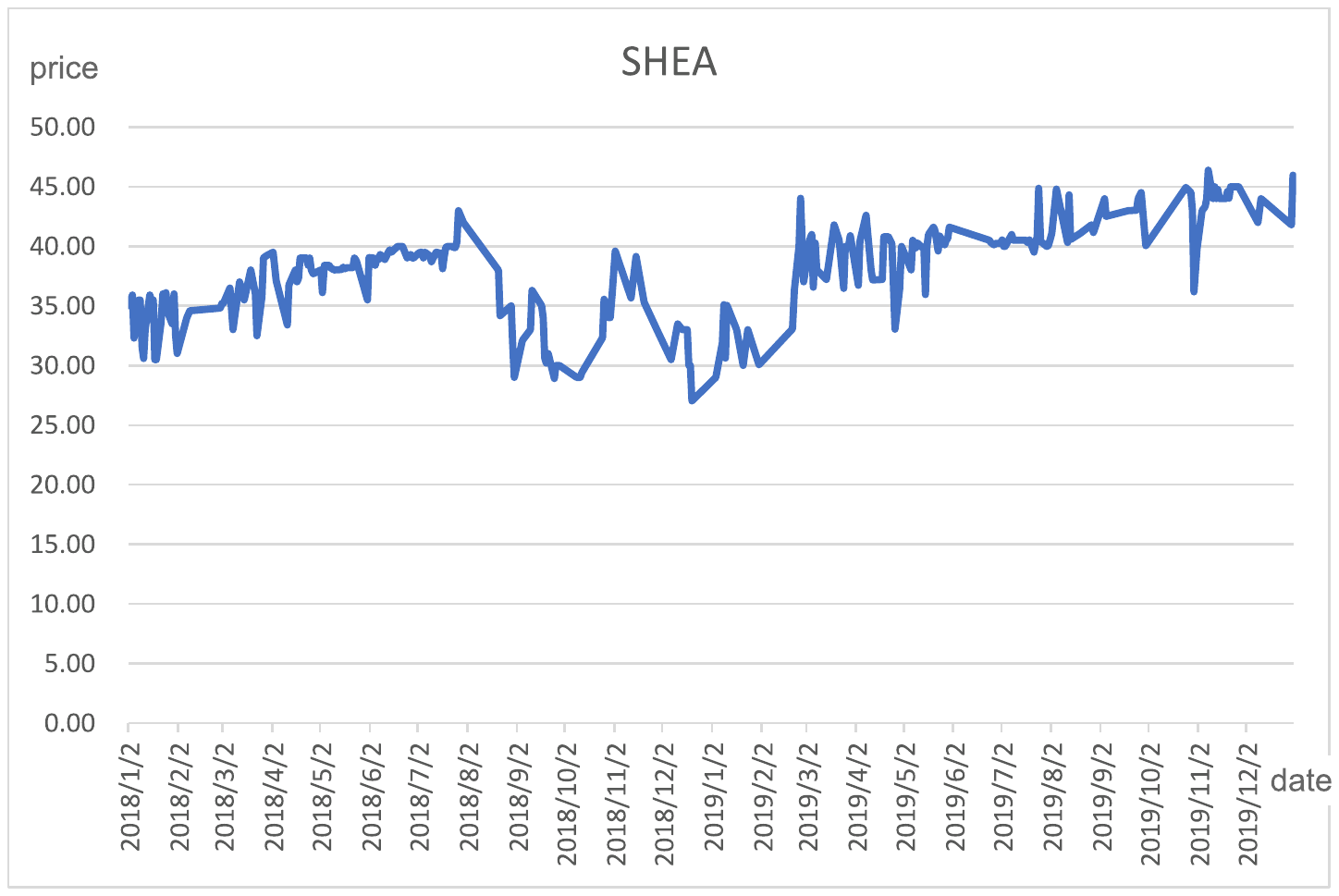}
\vskip-9.08368cm
	\caption{\small{Prize dynamics of SHEA at Shanghai Environment and Energy Exchange during year 2018-2019}}
	\label{price1819}
\end{figure}
\vskip-1.08368cm

\begin{center}
\begin{table}
\footnotesize
\label{table4}
\centering
\begin{tabular}{ccccccccccc}
\toprule
$Y_0$ & $\mu$ & $\sigma$ & $T$ &$P$ & $M$ &${\tilde P}$ & ${\tilde M}$ \\
\midrule
40.25 &0.0007 &0.0600 &  60 & 16.8 & 0.040 & 17.1 & 0.038\\
\bottomrule
\end{tabular} \caption{Parameters of the optimal stopping problem for analyzing the case of Wushashan power plant in 2019}
\end{table}
\end{center}
\vskip-1.18cm

\begin{figure}[htbp]
\vskip-1.38cm
\centering
\includegraphics[height=1.498 \textwidth,width=0.98\textwidth]{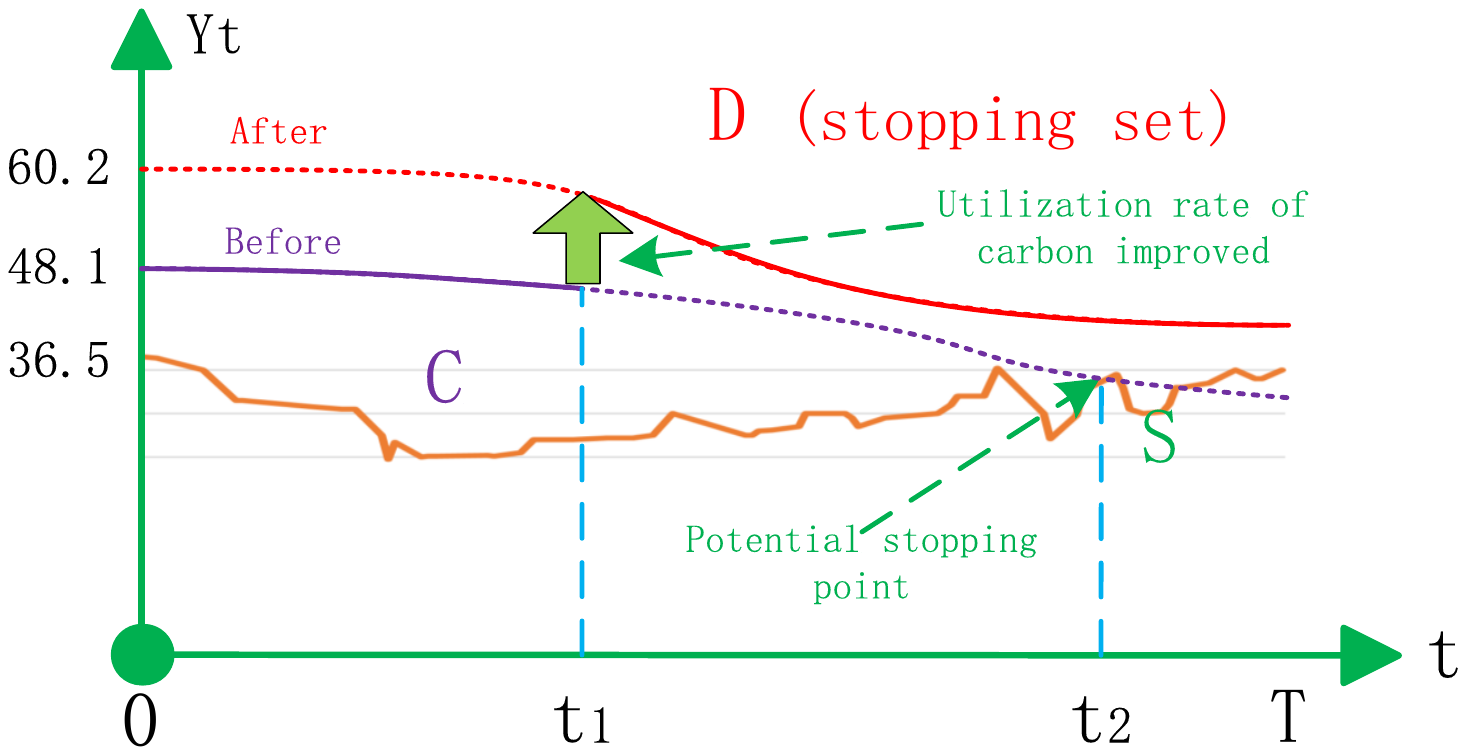}
\vskip-14.968668cm
	\caption{\small{Upward shift of optimal stopping boundary in the case of Wushashan 2017}}
	\label{tu2017}
\end{figure}

\begin{figure}[htbp]
\vskip-4.68cm
\centering
\includegraphics[height=1.498 \textwidth,width=0.98\textwidth]{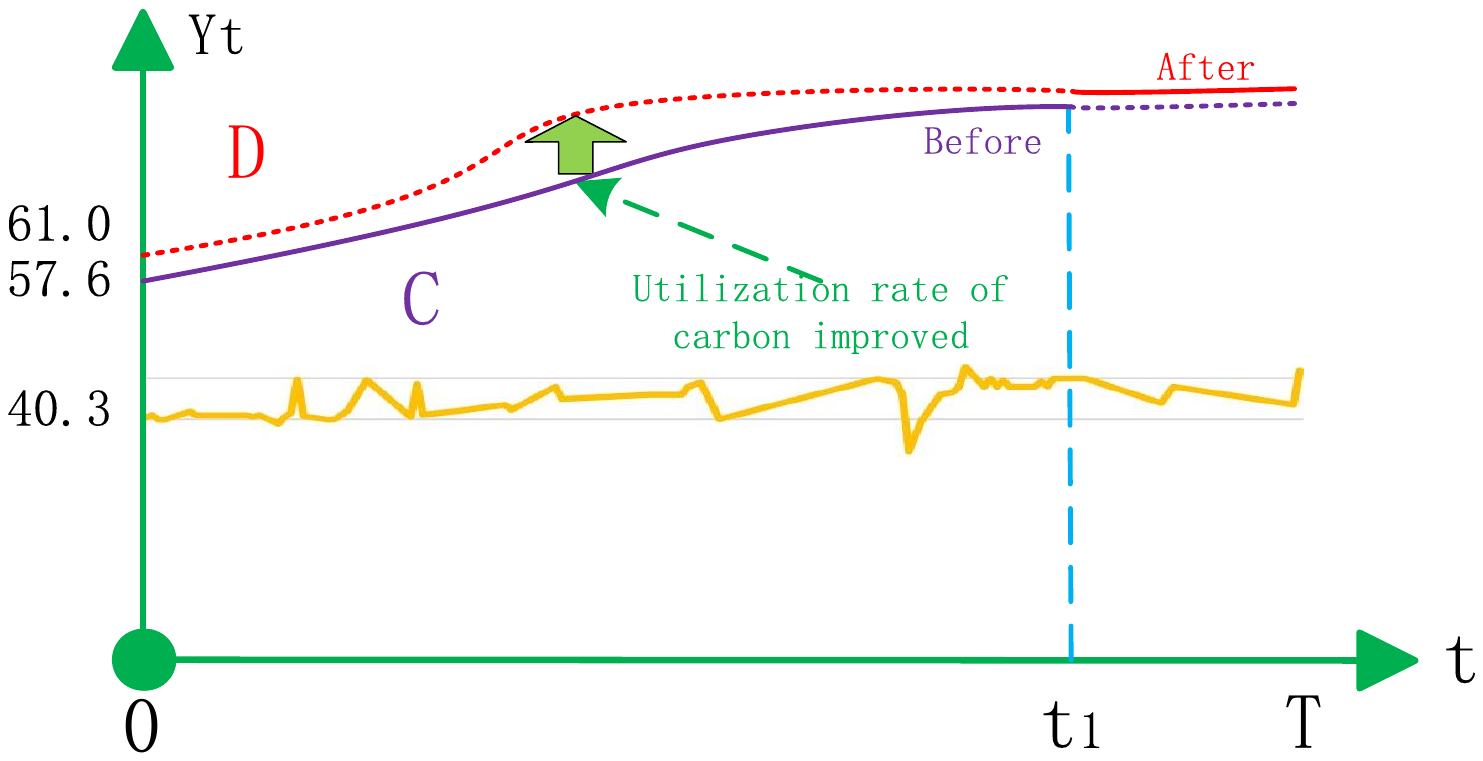}
\vskip-14.968668cm
	\caption{\small{Upward shift of optimal stopping boundary in the case of Wushashan 2019}}
	\label{tu2019}
\end{figure}

\noindent For the above two cases, we implement the backward recursive algorithm whose pseudocode is attached in the Appendix to solve the optimal stopping problem. As shown in Figure \ref{tu2017}, two optimal stopping boundaries are plotted by regressing the computational results with SSE less than 0.035 applying Matlab fitting tool, and the orange curve is plotted by connecting the real market prices. The lower purple curve denotes the boundary before technical upgrade, illuminated from the intersection of this curve and the market price curve (blue one, plotted with out-of-sample data), termination of business would be suggested at time $t_2$ in October this year (2017), as shown at the point marked by green 'S'. Fortunately, it is not the reality. At time $t_1$ in Aug. 2017, {\bf Technical upgrades pulled up the stopping boundary line} in time. Hence we find this purple curve half solid and half dash, and so is the red one above. This ascending happened at August (denoted by the green arrow in Figure \ref{tu2017}) and caused a fat rise of $P$ value, namely the averaged profit achievable at cost of one unit of carbon emission. Therefore, the stopping point 'S' is not actually encountered and the optimal stopping boundary afterwards is substituted by the solid red curve, which is well above the market price curve, hence the production activity remained running for all year 2017. Since then and for a long time, there were no conjunctures or challenges for Wushashan power plant. Figure \ref{tu2019} shows the case of 2019, as we mentioned for the results of investigation, mainly caused by developing the AGC and steam temperature intelligent control system, as well as the improvement of automatic control performance, Wushashan power plant again raised the utilization rate of carbon at Nov. 2019, marked as $t_1$ in Figure \ref{tu2019}. This upgrading provides upward shifting of the previous stopping boundary (purple one) to form a new stopping boundary (red one) although the previous boundary (before Nov. 2019) initially would have no chance to encounter the market price curve. Comparison between Figure \ref{tu2017} and Figure \ref{tu2019} may indicate a positive relation between sign of $\mu$ and the monotonicity of stopping boundary. However, suppose we consider the stopping boundary curves being overturned upside-down given a negative $\mu$ when the price of carbon emission allowance was downward, there were still enough margin above the market price curve (orange one) hence Wushashan power plant should well remain its operation. From the two cases of Wushashan power plant, we clearly see the seasonable technical upgrades may pull out the enterprise from probable dilemma.

\section{Surface of tipping benchmark}\label{outlook}
\begin{figure}[htbp]
\vskip-1.6868cm
\centering
\includegraphics[height=1.28 \textwidth,width=0.99\textwidth]{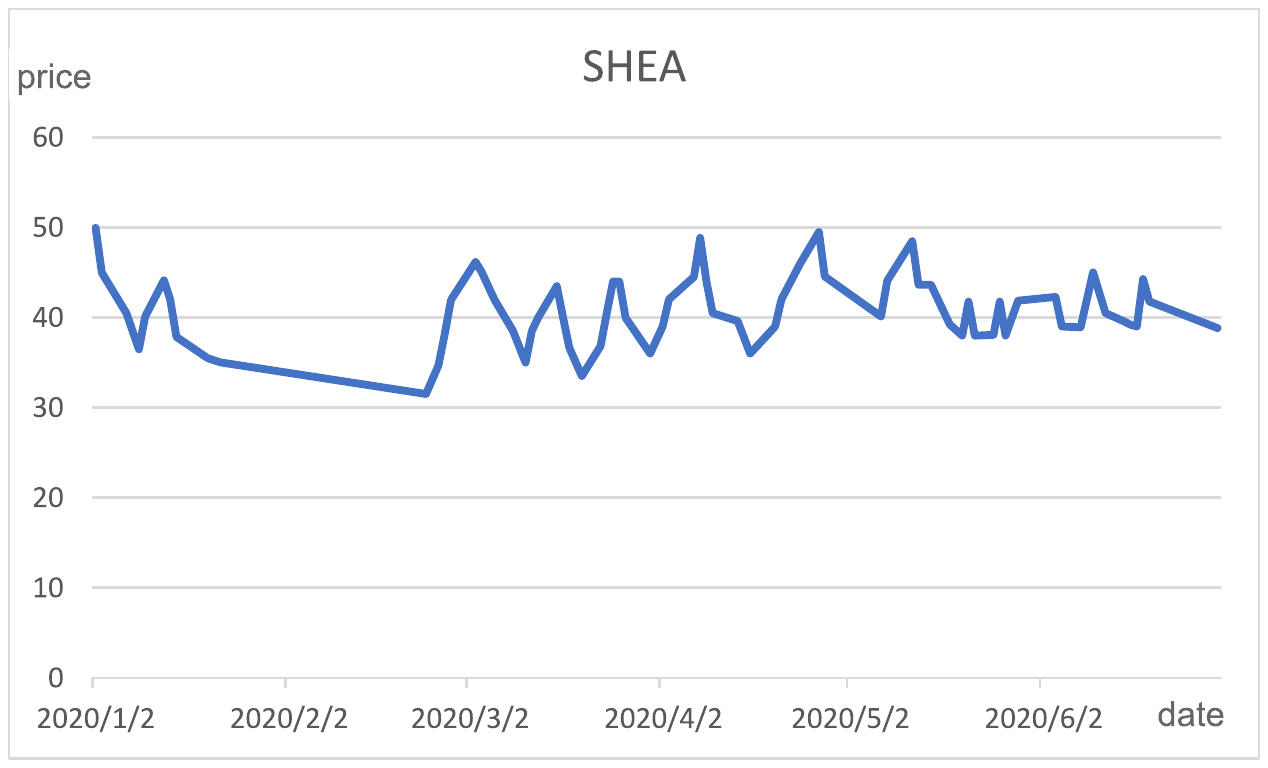}
\vskip-12.89368cm
	\caption{\small{Prize dynamics of SHEA at Shanghai Environment and Energy Exchange during first half year of 2020}}
	\label{shea2020}
\end{figure}

In the sequel we investigate the outlook of the survival environment of thermal power plants by presenting a diagram of the optimal stopping boundaries. Based on the data of first half year of 2020, shown in Figure \ref{shea2020} of SHEA at Shanghai Environment and Energy Exchange, we establish the price model \eqref{yt} with $\mu=-0.0014$ and $\sigma=0.0805$. Next, we consider the optimal stopping problem in \eqref{m} for different $P$ values (the averaged profit achievable at cost of a unit of carbon emission) and set $T=150$ for the year ahead (from Jul. 2020 to Jun. 2021) since the past one year has about $150$ trading days of SHEA.

\begin{figure}[htbp]
\vskip-8.38cm
\centering
 \hskip-1.0cm
\includegraphics[height=1.608 \textwidth,width=1.098\textwidth]{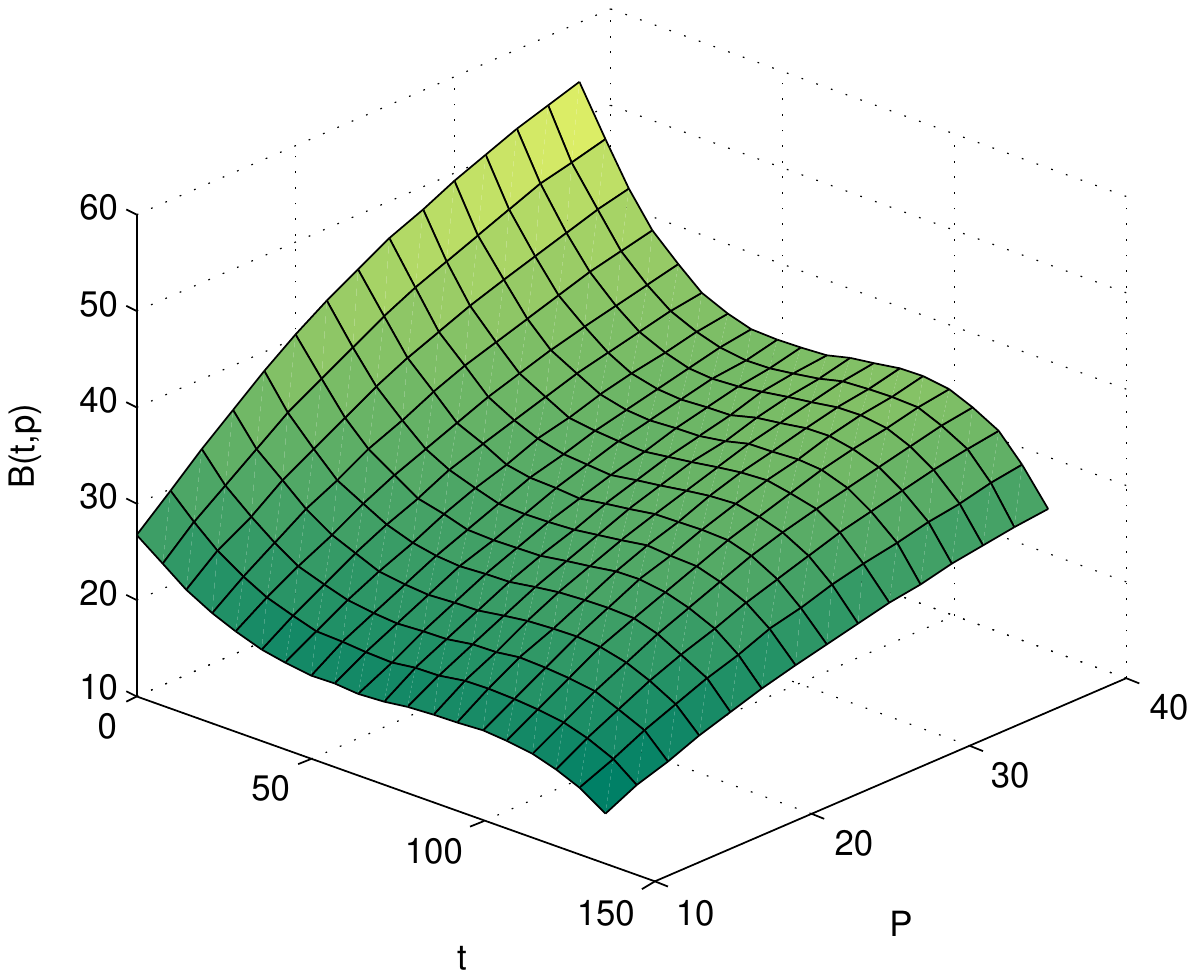}
\vskip-7.968668cm
	\caption{\small{Optimal stopping surface}}
	\label{last}
\end{figure}

\noindent Solving the optimal stopping problems for different $P$ values and collecting all boundary lines yield a surface in Figure \ref{last}. Inspecting this graph from a vertical angle, for any state $(t,p)$ that the $P$ value at time $t$ is $p$, we can find a unique point on the surface, denoted as $(t, p, B(t,p))$. Surviving or dying is illustrated by whether the carbon price is below or above $B(t,p)$.  From another angle, given carbon price $Y_t$ and at time $t$, a minimum value of $P$ is detected on the surface to satisfy the survival condition that $B(t,p)>Y_t$.
\\

\noindent It is obviously observed that, for any fixed $P\in[10,40]$, the boundary curve is decreasing in time $t\in[0,T]$ since the price trend of carbon emission allowance is downward as shown by Figure \ref{shea2020} regrading SHEA hence $\mu$ is negative. Besides, by raising the $P$ value, boundary curve is also lifted accordingly, which is consistent with the cases of
Wushashan power plant in 2017 and 2019. From another angle, considering the curve with increasing $P$ and fixed $t$, slope of the curve turns milder. This observation indicates that, it gets comparatively tougher to uplift the optimal stopping boundaries when $P$ is achieving a higher value. Unfortunately, we feel even more concerned when we realize that, this law of diminishing marginal utility is not isolate, it has already existed in the process of raising the $P$ value by technical upgrades.\\

\noindent Comparing with Figure \ref{shea2020}, we seriously concern that, the survival environment of thermal power industry is tough since the high price of carbon emission rights allowance is always challenging. Though the demands of capacity declined due to the outbreak of COVID-19 epidemic, the price remains in a high level about 40 CNY, and there is no sign of its cooling down after the full resumption of work in Chinese mainland. According to Figure \ref{last}, for those thermal power plants with their $P$ value under 40, it is very likely for the market price curve to break through their feeble boundaries. That is, those plants will be nearly inevitably trapped in a state that, suspending or halting production overwhelm other options. Thereafter, overall shutdown may come in the near future, like the ending process in the case of Shajiao B power plant.

\section{Summary and conclusions}\label{conclu}
\noindent Pushed by the great cause of carbon neutrality, the survival environment of thermal power industry becomes severer as carbon regulation turns stricter. Intuitionally, two directions towards survival are feasible, one is to reduce absolute quantity of carbon emission as the rule makers wish, another is to boost up the output value with the same consumption of carbon emission. See from several cases we presented above, both approaches can be achieved by technical upgrades and reflected on a key variable $P$ value investigated in the above text, which counts for the averaged net profit generated by consuming per unit of carbon emission.
\\

\noindent To illustrate the relation between $P$ value and the viability of thermal power plants, an optimal stopping model is designed to exactly reflect their dynamic connection. Once the model suggests a stopping, it means that suspending or halting of production right at that circumstance is the optimal choice compared with maintaining operation, this decision is made with respect to the maximization of ultimate profit. Without any effective and immediate remedial measures pulling up it to exceed the stopping boundary, shutdown follows its failure of operation, as shown by the first case, namely the shutdown of Shajiao B power plant in 2019. Alternatively, technical improvements to enhance the utilization rate of carbon emission is the main countermeasure, which reverses this dilemma by placing the optimal stopping boundaries at a safe position well above the market price curve of carbon emission allowance, as shown by the cases of Wushashan power plant in 2017 and 2019.
\\

\noindent Applying the optimal stopping model based on the market data of 2020, we present the outlook of survival environment of thermal power industry in the near future. By displaying the tipping benchmark of production halts, Figure \ref{last} answers to the question that at least how much is the $P$ value of a thermal power plant to survive under the pressure of high price of carbon emission allowance. The most important observation is, by increasing the $P$ value, optimal stopping boundaries will rise accordingly. Besides, we find that, when the price of carbon emission allowance is in a upward (downward) trend, the boundary curves are also increasing (decreasing) in $t\in[0,T]$.
\\

\noindent Given the price of carbon emission allowance continues the existed trend and the SHEA price maintain the level around 45 CNY, thermal power plants are compelled to ensure their $P$ value (namely the averaged net profit generated by consuming per unit of carbon emission) well above 40 CNY. Associated with the stopping surface in Figure \ref{last}, double pressures of diminishing marginal utility implies that, it is increasingly difficult to uplift the $P$ value by technical upgrades as well as to uplift the boundary curves by prising $P$ values. Along this trend, the vivosphere of thermal power plants is shrinking. The ultimate savior for them must be the government conversely, although it will not loosen the carbon regulation by denying the previous efforts. Several solutions are feasible, for one thing, government is able to suppress the carbon price from surging (for instance, SHEA price should be controlled during 35 to 45 CNY), for another, if the above intervention is blamed for disturbing the overall cause of carbon reduction, government can still take its old way by further increasing subsidies for thermal power plants which will also raise their $P$ values immediately. Anyway, those begging for survival will finally be weeded out from the system of laying off the last even when the government offers more subsidies to this industry. Without big generator sets, traditional thermal power plants might be forced to transform towards new energy like solar or photovoltaic power. For them, solar aided coal based power generation is a potential alternative, or to share part of the components with other power generation is a compromise approach. Policy design should be appropriately biased to guide this transformation by resetting the carbon emission tax. Indeed, though the carbon emission tax is not considered separately, it been has already taken account when we measure the profit $P$. Further research may consider the carbon emission tax as an independent variable in a more profound but also more complex model.

\section*{Appendix}
\begin{algorithm}[h!]
\caption{Algorithm to solve the optimal stopping problem}\label{algo}
\begin{algorithmic}[1]
\Require
Time boundary $T$,  take-profit level $K$, drift factor $\mu$, volatility $\sigma$.
\Ensure
$G(t,y)$ and $V(t,y)$ for each $t\in\{t_0,t_1,\ldots,t_n\}$ and $y\in\{y_0,y_1,\ldots,y_m\}$, $b(t)$ for each $t\in\{t_0,t_1,\ldots,t_n\}$.
\ForAll{$t\in\{t_0,t_1,\ldots,t_n\}$}
\ForAll{$y\in\{y_0,y_1,\ldots,y_m\}$}
\State calculate $G(t,y)$ according to \eqref{g2};
\EndFor
\EndFor
\For{each $t\in\{t_n,t_{n-1},\ldots,0\}$ {\bf backwardly}}
\For{each $y\in\{y_0,y_1,\ldots,y_m\}$ {\bf increasingly}}
\For{each sample $Y_{t_k}$ given $Y_{t_{k-1}}=y$}
\If{$V(t_k,Y_{t_k})>G(t_k,Y_{t_k})$}
\State obtain a sample $V(t_{k-1},y)=G(t_k,Y_{t_k})$;
\Else[$V(t_k,Y_{t_k})\leq G(t_k,Y_{t_k})$]
\State obtain a sample $V(t_{k-1},y)=V(t_k,Y_{t_k})$;
\EndIf
\State average the samples of $V(t_{k-1},y)$ to get $V(t_{k-1},y)$;
\EndFor
\EndFor
\EndFor
\For{each $t\in\{t_n,t_{n-1},\ldots,0\}$}
\For{each $y\in\{y_0,y_1,\ldots,y_m\}$ {\bf increasingly}}
\If{$V(t,y)=G(t,y)$}
\State exit for loop of $y$;
\EndIf
\EndFor
\State let $b(t)=y$;
\EndFor
\end{algorithmic}
\end{algorithm}
$
$
$
$
$
$
$
$
$
$\\
$
$
$
$
$
$\\
$
$
$
$
$
$

\end{document}